\documentclass[pss]{wiley2sp} 
\usepackage[numbers,sort&compress]{natbib}
\usepackage{amsmath}
\usepackage{units}

\usepackage[bookmarks=true,letterpaper=true,colorlinks=true,urlcolor=blue,linkcolor=blue,citecolor=blue]{hyperref} 
\def\copyrightmark{{\copyrightmarkfont}}%
\journalname{submitted to physica status solidi}

\begin{document}

\title{
Designing multifunctional chemical sensors using Ni and Cu doped carbon nanotubes
}

\titlerunning{Designing multifunctional sensors using doped carbon nanotubes}

\author{%
  D. J. Mowbray\textsuperscript{\Ast,\textsf{\bfseries 1,2}},
  J. M. Garc{\'{\i}}a-Lastra\textsuperscript{\textsf{\bfseries 1,2}},
  K. S. Thygesen\textsuperscript{\textsf{\bfseries 2}},
  A. Rubio\textsuperscript{\textsf{\bfseries 1,3}},
  K. W. Jacobsen\textsuperscript{\textsf{\bfseries 2}}}

\authorrunning{D. J. Mowbray et al.}

\mail{e-mail
  \textsf{duncan\_mowbray@ehu.es}, Phone:
  +34-943-01-5366, Fax: +34-943-01-8390}

\institute{%
  \textsuperscript{1}\,Nano-Bio Spectroscopy Group and ETSF Scientific Development Centre, Dpto.~F{\'{\i}}sica de Materiales, Universidad del Pa{\'{\i}}s Vasco, Centro de F{\'{\i}}sica de Materiales CSIC-UPV/EHU-MPC and DIPC, Av.~Tolosa 72, E-20018 San Sebasti{\'{a}}n, Spain\\
  \textsuperscript{2}\,Center for Atomic-scale Materials Design, Department of Physics,
  Technical University of Denmark, DK-2800 Kgs.~Lyngby, Denmark\\
  \textsuperscript{3}\,Fritz-Haber-Institut der Max-Planck-Gesellschaft, Berlin, Germany}

\received{\today}

\keywords{nanosensing, metal-doping, nanotubes, CO, NH$_{\text{3}}$.}

\abstract{%
%
%
%
\abstcol{%
We demonstrate a ``bottom up'' approach to the computational design of a multifunctional chemical sensor.  General techniques are employed for describing the adsorption coverage and resistance properties of the sensor based on density functional theory (DFT) and non-equi-}{librium Green's function methodologies (NEGF), respectively. Specifically, we show how Ni and Cu doped metallic (6,6) single-walled carbon nanotubes (SWNTs) may work as effective multifunctional sensors for both CO and NH$_{\text{3}}$.
}
}

%
%

\maketitle   

\section{Introduction}

\begin{figure}[tbp]
\centering
\includegraphics*[width=\columnwidth]{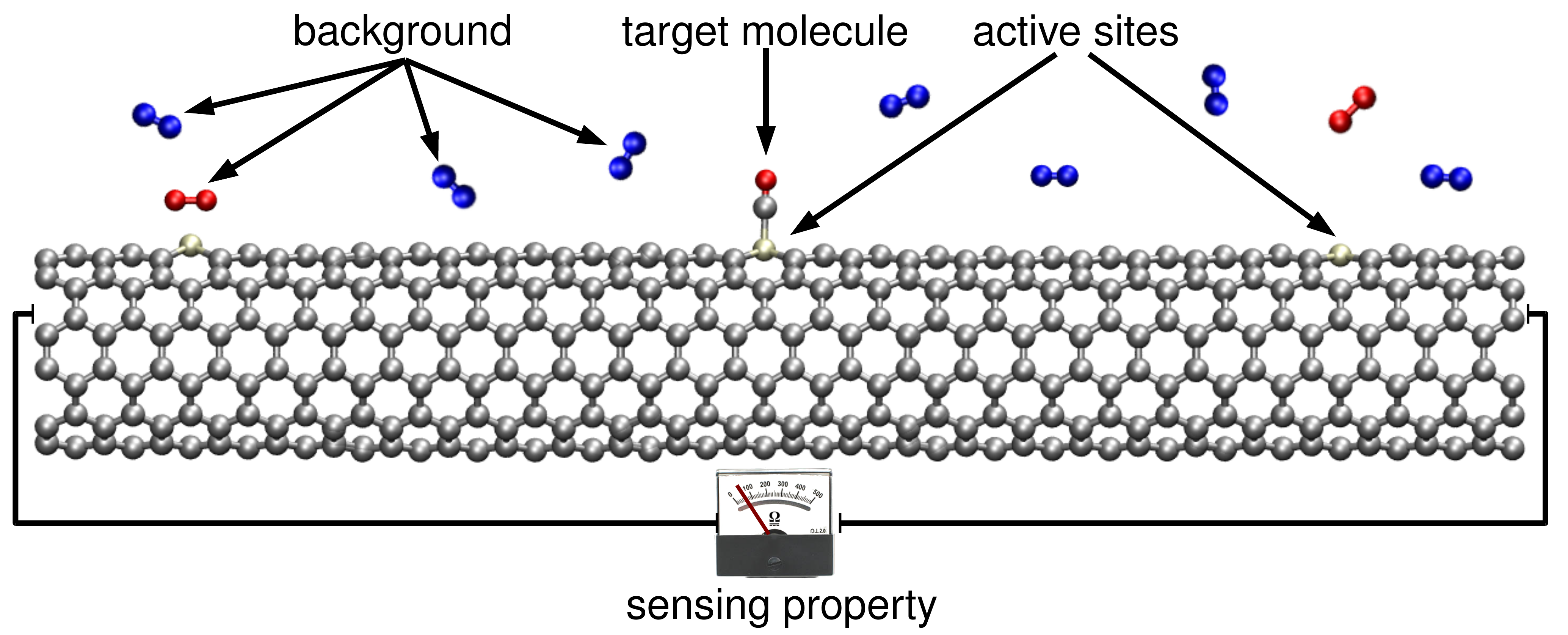}
\caption{Schematic of a chemical sensor consisting of active sites (metal dopants in a (6,6) carbon nanotube), a target molecule (CO), a background (atmospheric air), and a sensing property (resistance).
}
\label{Fig1}
\end{figure}


Detecting specific chemical species at small concentrations 
is of fundamental importance for many industrial/scientific processes, medical applications, and environmental monitoring \cite{gas_sensing}.  Nanostructured materials are ideally suited for sensor applications because of their large surface to volume ratio, making them sensitive to the
adsorption of individual molecules. Specifically, single-walled carbon nanotubes (SWNTs) \cite{cnt_review} work remarkably well as detectors for small gas molecules, as demonstrated for both individual
SWNTs \cite{kong,collins,hierold,villalpando,rocha,Brahim} and SWNT networks \cite{morgan,cnt_networks}. 
Previous studies have shown that SWNTs are highly sensitive to most molecules upon functionalization \cite{Fagan,Yagi,Yang,Chan,Yeung,Vo,Furst,Juanma,Krasheninnikov}.  However, the difficulty is determining which specific molecules are present.  In this study we show how changing the functionalization provides ``another handle'' for differentiating the SWNTs response of different gases/molecules.

Recent experimental advances now make the controlled doping of chirality selected SWNTs with metal atoms a possibility.  Specifically, these include (1) photomoluminesence, Raman and XAS techniques for measuring the fraction of various SWNT chiralities in an enriched sample \cite{Kramberger07PRB,Ayala09PRB,DeBlauwe2010}; (2) the separation of SWNT samples by chirality using DNA wrapping \cite{Zheng03NM,Zheng03S,TU09N,Li07JOTACS}, chromatographic separation
\cite{Li07JOTACS,Fagan07JOTACS} and Density Gradient Ultracentrifugation
(DGU)\cite{Arnold06NN}; (3) SWNT resonators for measuring individual atoms of a metal vapour which adsorb on a SWNT \cite{Bachtold,ZettlMassSensor}; and (4) aberration corrected low energy ($<$50 keV) transmission electron microscopy (TEM) \cite{Chuanhong}.  The latter provides control over the formation of defects \emph{in situ} by adjusting the energy of the electron beam above and below the threshold energy for defect formation at a specific location in the SWNT.  These methods provide such a high level of control that it is now possible for experimentalists to take a specific SWNT chirality and dope the structure with individual metal atoms at a specified location.

At the same time, theorists are now able to embrace a ``bottom up'' approach to the design of nanosensors, harnessing the thermodynamics of self-assembly to find useful sensing systems \emph{in silico}.  With recent advances in both computational power and methodologies, theorists can now efficiently and accurately screen hundreds of candidate sensor designs using a combination of density functional theory (DFT) for energetics of adsorption and stability, and non-equilibrium Green's function (NEGF) methodologies for the electrical response \cite{Juanma2}.


In general, any nanosensing system consists of the following
four main components: (1) a ``target molecule'' to be detected, (2) an
``active site'' where the target molecule may adsorb on the sensor,
(3) a property of the sensor which changes upon adsorption of the
target molecule, and (4) a ``background'' of adsorbing molecules which
make up the background signal, as depicted in Fig.~\ref{Fig1}.  For the sensor to be effective, the active site must be designed so that adsorption of the target molecule in the presence of the background is sufficient to change the sensing property.  

\begin{figure}[tbp]
  \centering
  \includegraphics*[width=0.855\columnwidth]{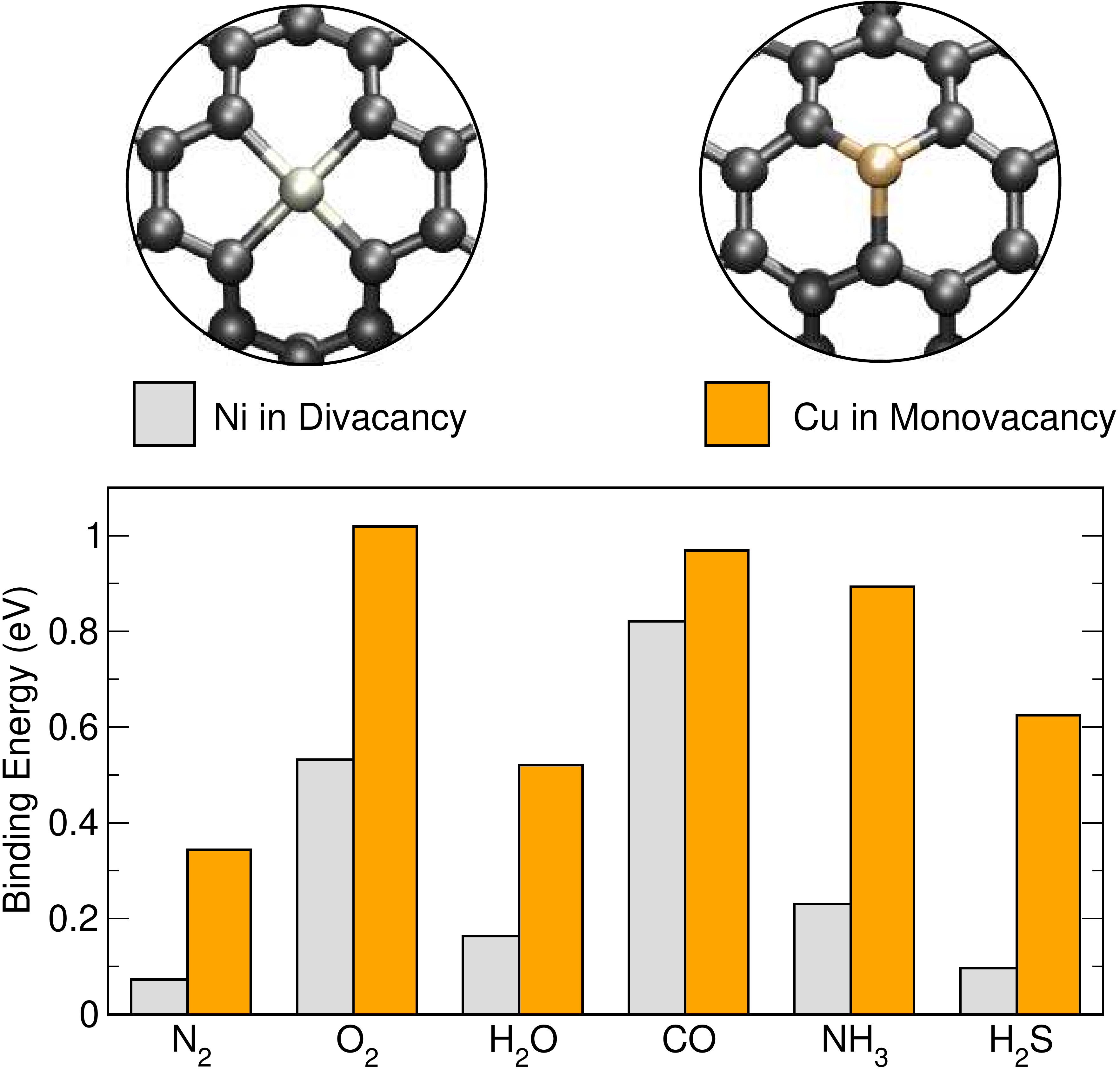}%
  \caption[]{Binding energies in eV for N$_{\text{2}}$, O$_{\text{2}}$, H$_{\text{2}}$O, CO, NH$_{\text{3}}$, and H$_{\text{2}}$S on a Ni atom occupying a divacancy site (grey) and on a Cu atom occupying a monovacancy site (orange) in a (6,6) carbon nanotube.  Schematics of the two active sites are shown above.
}
    \label{Fig2}
\end{figure}

\begin{figure}[tbp]
\centering
\includegraphics*[width=\linewidth]{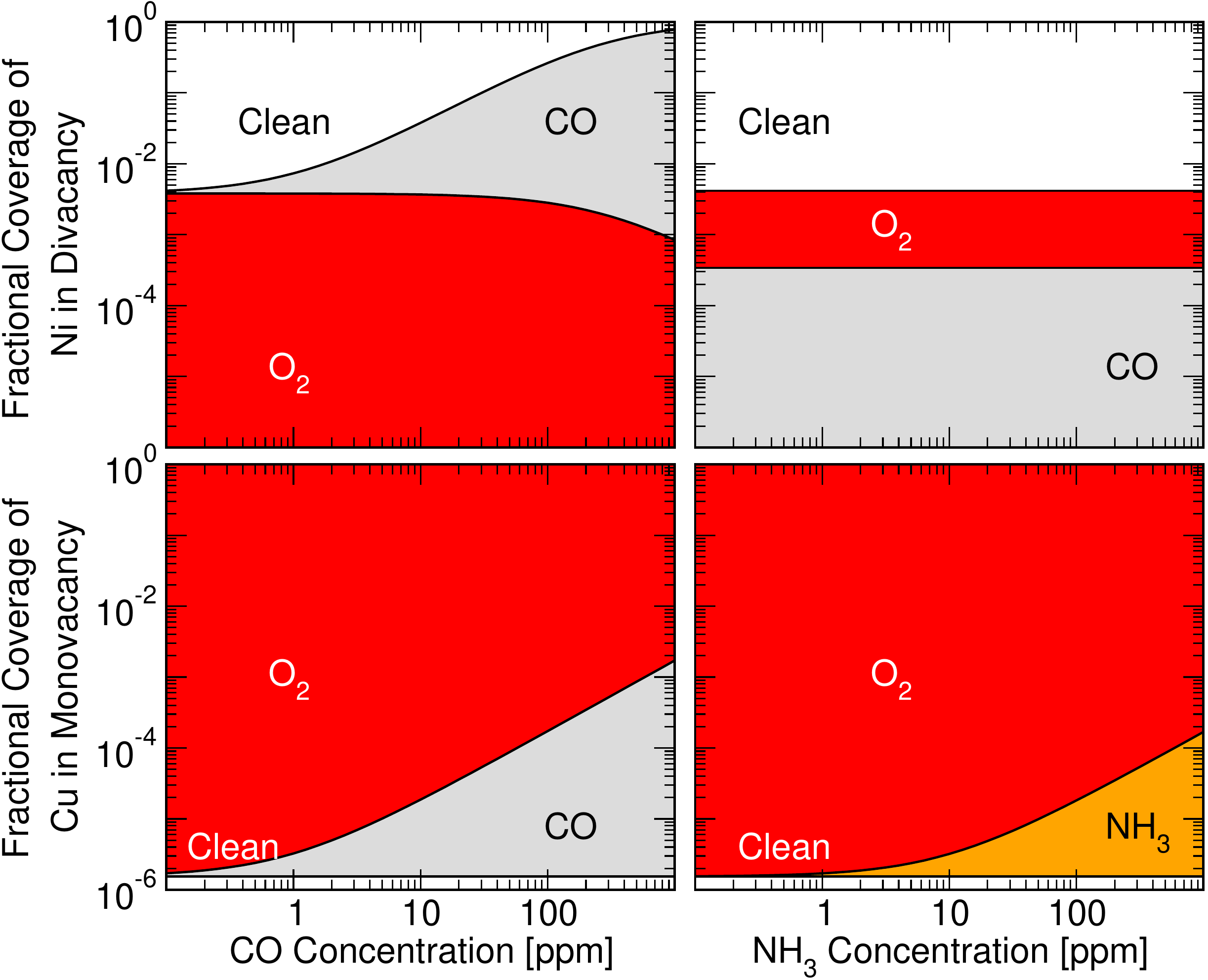}
\caption{%
 Fractional coverage in thermal equilibrium of Ni in a divacancy (top) and Cu in a monovacancy (bottom) versus CO concentration (left) and NH$_{\text{3}}$ concentration (right) in a background of air at room temperature and 1 bar of pressure.}
\label{Fig3}
\end{figure}

In this study we shall consider Ni and Cu doped metallic (6,6) SWNTs as our active sites, specifically when Ni replaces two C atoms in a divacancy structure, and Cu replaces a single C atoms in the monovacancy structure, as shown schematically in Fig.~\ref{Fig2}.  Of the two possible orientations for a divacancy in an armchair SWNT, we here consider the one in which Ni adsorption has been shown to be more favourable. These two active sites have previously been shown to be both stable, and somewhat resistant to oxidation under standard atmospheric pressure and room temperature \cite{Juanma2}.  As such, we shall consider the background to have a standard atmospheric composition with an absolute humidity of 4\%, shown in Table \ref{Table1}.  

\begin{table}
\centering
\caption{Equilibrium atmospheric concentrations at 4\%  absolute humidity.}\label{Table1}
\begin{tabular}[tbhp]{crcr}
\hline
Background & Concentration & Target & Concentration\\\hline
N$_{\text{2}}$ & 74.96\%&CO &    96.00 ppb\\ 
O$_{\text{2}}$ & 20.11\%&NH$_{\text{3}}$&  16.32 ppb\\
H$_{\text{2}}$O& 4.00\%& H$_{\text{2}}$S& 0.96 ppb \\
\hline
\end{tabular}
\end{table}

\section{Methodology}

All DFT calculations have employed the PBE exchange correlation
(xc)-functional \cite{PBE}. Adsorption energies and structural
minimization calculations have been performed using the real-space
projector-augmented-wave method DFT code \textsc{gpaw} \cite{GPAW}, with
a grid spacing of 0.2 \AA, taking spin polarization into account.  
NEGF transport calculations \cite{benchmark} have employed the electronic Hamiltonian obtained from the DFT code \textsc{siesta} \cite{SIESTA}, with a double zeta polarized (DZP) localized basis set.   

Vacancies in the (6,6) metallic carbon nanotube have been modelled
using six repeated minimal unit cells, with a supercell of 15 \AA\
$\times$ 15 \AA\ $\times$ 14.622 \AA.  Such a large supercell was
necessary to both minimize interactions between defects and adsorbates
and to ensure the Hamiltonian of the nanotube layers adjacent to the
cell boundaries was converged to within 0.1 eV of that for the
pristine nanotube. For such a large supercell, a \(\Gamma\) point
calculation was found to be sufficient to describe the nanotube's
periodicity.   

Binding energies $E_{\text{bind}}$ are defined as the difference in total energies between a gas phase target/background molecule with a clean active site and a target/background molecule adsorbed on an active site.  As seen in Fig.~\ref{Fig2}, CO has a strong binding energy to both active sites, while NH$_{\text{3}}$ binds only to Cu in a monovacancy site.  We shall use this in our design of a multifunctional sensor for both CO and NH$_{\text{3}}$.  

\section{Results \& Discussion}

\begin{figure}[tbp]
\centering
\includegraphics*[width=\linewidth]{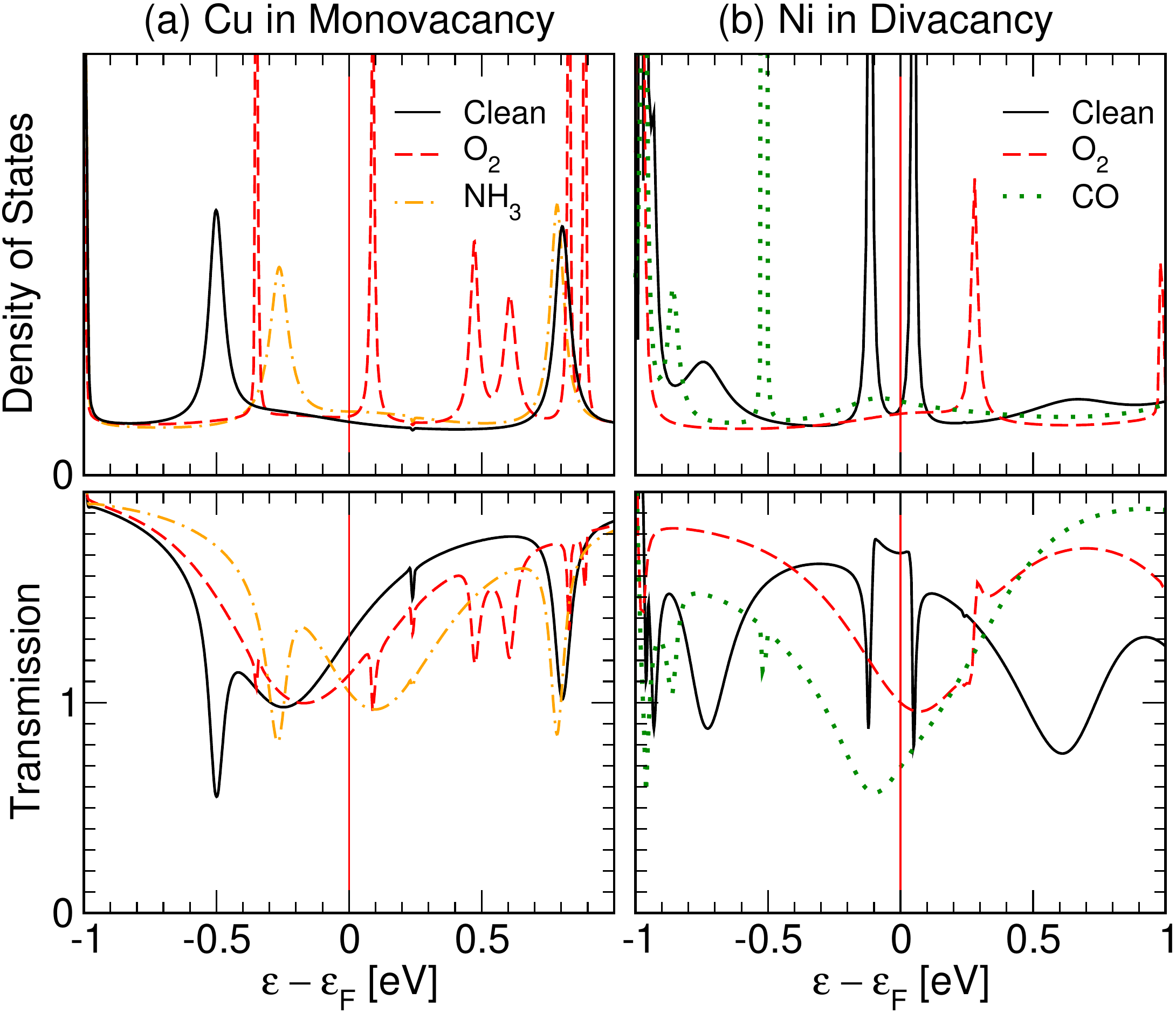}
\caption{%
Density of states (top) and transmission function (bottom) versus energy \(\varepsilon\) relative to the Fermi level \(\varepsilon_{\rm F}\) in eV for clean (solid line; black), O$_{\text{2}}$ occupied (dashed line; red), NH$_{\text{3}}$ occupied (dash-dotted line; orange), and CO occupied (dotted line; green) (a) Cu in monovacancy and (b) Ni in divacancy II of a (6,6) carbon nanotube.  
}
\label{Fig4}
\end{figure}

For a sensor containing hundreds, if not thousands of sites, the fractional coverage of sites is reasonably described by the fractional coverage in thermodynamic equilibrium $\Theta$ at standard temperature and pressure in terms of the target molecule concentration.  Specifically for a sufficiently large ensemble of active sites, the fractional coverage of sites by a given adsorbate $X$ in a background set of molecules $\mathcal{B}$ at equilibrium is given by
\begin{eqnarray}
\Theta[X] &=& \frac{K[X] C[X]}{1 + \sum_{Y \in \mathcal{B}} K[Y] C[Y]},
\end{eqnarray}
where $K = \nicefrac{k_+}{k_-}$ is the ratio of forward and backward
rate constants for the adsorption reaction. This may be expressed as 
\begin{eqnarray}
K[X] &=& \exp \left[\frac{ E_{\text{bind}}[X] - T S_{\text{gas}}[X]}{k_B T}\right],
\end{eqnarray}
where $S_{\text{gas}}[X]$ is the gas phase entropy of species $X$ \cite{CRCHandbook}, $k_B$ is Boltzmann's constant and $T$ is the temperature.

In Fig.~\ref{Fig3} we see the coverages of both Ni and Cu active sites as a function of CO and NH$_{\text{3}}$ concentration.  Clearly, both Cu and Ni active sites will be sensitive to the adsorption of CO.  On the other hand, the coverage of the Cu active sites is sensitive to NH$_{\text{3}}$ concentration, while Ni active sites are not.  This suggests that by combining the response of Cu and Ni active sites, we may obtain a multifunctional sensor for both CO and NH$_{\text{3}}$.

\begin{figure}[tbp]
\centering
\includegraphics*[width=\linewidth]{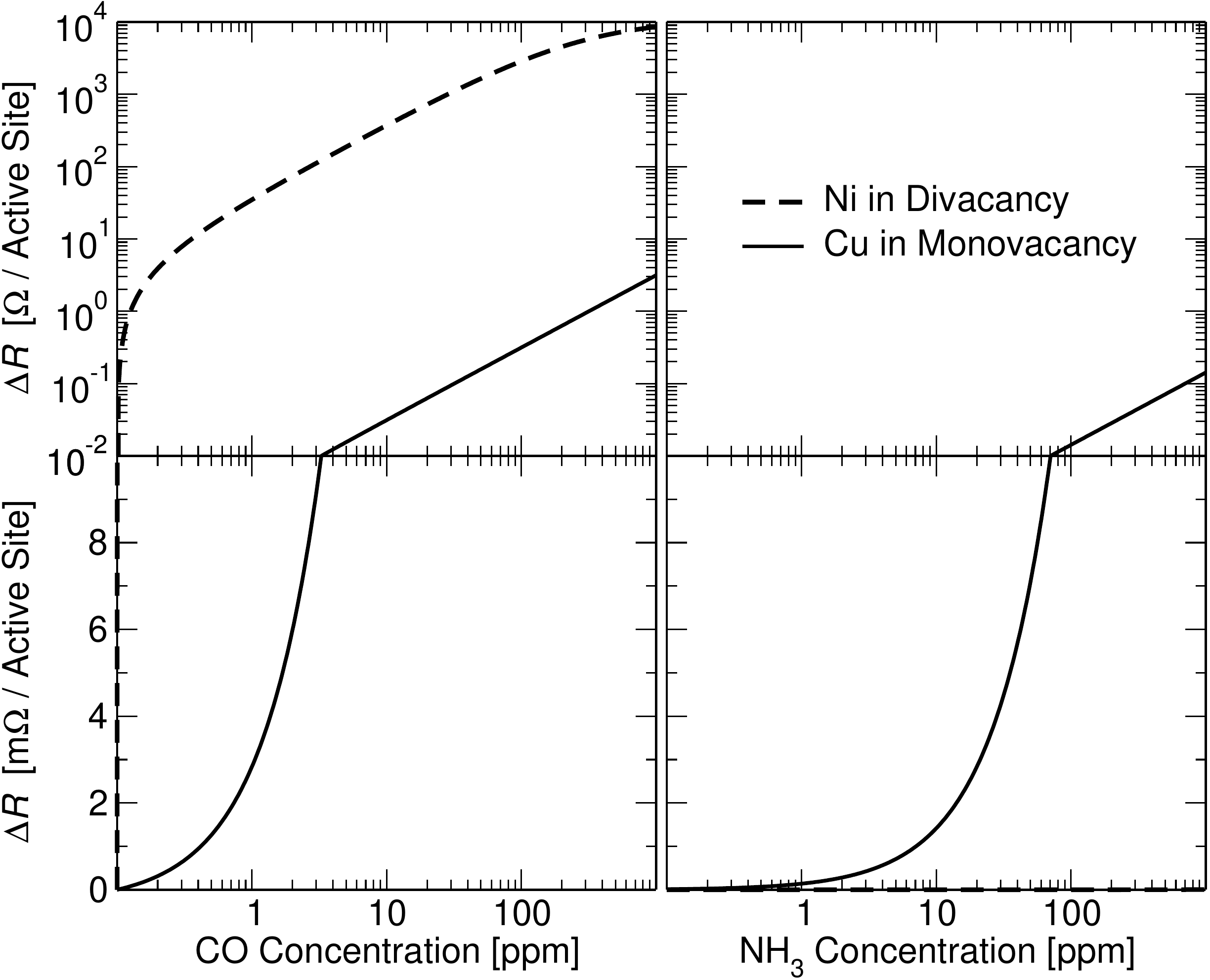}
\caption{%
Change in resistance \(\Delta R\) in $\unit{\Omega}$ (top) and m$\unit{\Omega}$ (bottom) per active site versus CO (left) and NH$_{\text{3}}$ (right) concentration in ppm for Ni in a divacancy II (dashed line) and Cu in a monovacancy (solid line) of a (6,6) carbon nanotube.  The reference concentrations of CO and NH$_{\text{3}}$ are 0.1 and 0.01 ppm, respectively, in a background of air at room temperature and 1 bar of pressure.}
\label{Fig5}
\end{figure}

However, although molecules may adsorb on the active sites, the sensing property of the SWNT resistance must change significantly for the sensor to be useful.  In Fig.~\ref{Fig4} we show both the density of states (DOS) and transmission probability T($\varepsilon$)  for an electron of a given energy $\varepsilon$ to travel past an active site as calculated within the NEGF method.  Since the energy levels of a given molecule may be thought of as its electronic ``finger print'',  so long as there is sufficient binding between the molecule and active site, this ``finger print'' is evidenced in the DOS of the system.  Furthermore, the peaks of the DOS due to the molecule tend to form Fano anti-resonances in the transmission, as electrons are scattered off these states (\emph{c.f.} Fig.~\ref{Fig4}).  In this way, any adsorbed molecule tends to leave its ``finger print'' in the transmission probability through an active site.  For this reason, conductance/resistance is generally an effective sensing property.

Finally, we may estimate the average change in resistance $\Delta R$ of an active site as a function of target molecule concentration.  As discussed in Ref.~\cite{Juanma2}, this change in resistance is reasonably well described by
\begin{eqnarray}\label{eq.totalres}
\Delta R &\approx& \sum_{X} R_s(X)(\Theta[X,C] - \Theta[X,C_0]), 
\end{eqnarray}
where $C$ is the concentration, $C_0$ is the concentration at standard temperature and pressure, $R_s(X)=G_0({\text{1}}/\text{T}(\varepsilon_F,X)-{\text{1}}/({\text{2}})$, $G_0 \equiv \text{2} e^{\text{2}}/h$ is the quantum of conductance, and T$(\varepsilon_F,X)$ is the transmission probability at the Fermi level through an active site with species $X$ adsorbed.  

In Fig.~\ref{Fig5} we show $\Delta R$ for a single active site (Ni or Cu) as a function of target molecule concentration (CO or NH$_{\text{3}}$).  Keeping in mind that a change in concentration from 1 ppm to 50 ppm amounts to a change from allowable to toxic concentrations for both CO and NH$_{\text{3}}$, we see that this sensor design should work effectively for both target molecules.

\section{Conclusions}

In conclusion, we have demonstrated \emph{in silico} how a combined Cu and Ni doped metallic SWNT device may work effectively as a multifunctional sensor for both CO and NH$_{\text{3}}$.  The methodology employed is applicable to other sensor designs and different environments.  Also, by varying the metal dopant, we obtain ``another handle'' for tuning the sensitivity of our sensor.

\begin{acknowledgement}
The authors acknowledge financial support from 
Spanish MEC (FIS2007-65702-C02-01),
ACI-Promociona (ACI2009-1036),
``Grupos Consolidados UPV/EHU del Gobierno Vasco'' (IT-319-07), the European Union through the FP7 e-I3 ETSF (Contract Number 211956), and THEMA (Contract Number 228539) projects.  We acknolege support by the Barcelona Supercomputing Center, ``Red Espa{\~{n}}ola de
Supercomputaci{\'{o}}n'', ARINA, NABIIT and the Danish Center for Scientific
Computing. The Center for Atomic-scale Materials Design (CAMD) is
sponsored by the Lundbeck Foundation.  \end{acknowledgement}

%

\begin{thebibliography}{10}
\bibitem{gas_sensing}
\newblock{}{\emph{Gas Sensing Materials, MRS Bull.}}, {}{vol.~24} ({}{1999}).
\bibitem{cnt_review}
{}{J.~C. Chalier}, {}{X.~Blase}, and {}{S.~Roche}, \newblock{}{``Electronic and
  transport properties of nanotubes''}, \newblock{}{Rev. Mod. Phys.}
  {}{\textbf{79}}(2), {}{677} ({}{May 2007}), \newblock
  doi:\href{http://dx.doi.org/10.1103/RevModPhys.79.677}{10.1103/RevModPhys.79%
.677}.
\bibitem{kong}
{}{J.~Kong}, {}{N.~R. Franklin}, {}{C.~Zhou}, {}{M.~G. Chapline}, {}{S.~Peng},
  {}{K.~Cho}, and {}{H.~Dai}, \newblock{}{``Nanotube molecular wires as
  chemical sensors''}, \newblock{}{Science} {}{\textbf{287}}(5453), {}{622}
  ({}{Jan. 2000}), \newblock
  doi:\href{http://dx.doi.org/10.1126/science.287.5453.622}{10.1126/science.28%
7.5453.622}.
\bibitem{collins}
{}{P.~G. Collins}, {}{K.~Bradley}, {}{M.~Ishigami}, and {}{A.~Zettl},
  \newblock{}{``Extreme oxygen sensitivity of electronic properties of carbon
  nanotubes''}, \newblock{}{Science} {}{\textbf{287}}(5459), {}{1801} ({}{Mar.
  2000}), \newblock
  doi:\href{http://dx.doi.org/10.1126/science.287.5459.1801}{10.1126/science.2%
87.5459.1801}.
\bibitem{hierold}
{}{C.~Hierold}, \newblock{}{\emph{Carbon Nanotube Devices: Properties,
  Modeling, Integration and Applications}} ({}{Wiley-VCH}, Weinheim, {}{2008}).
\bibitem{villalpando}
{}{F.~Villalpando-P{\'{a}}ez}, {}{A.~H. Romero}, {}{E.~Mu{\~{n}}oz-Sandoval},
  {}{L.~M. Mart{\'{\i}}nez}, {}{H.~Terrones}, and {}{M.~Terrones},
  \newblock{}{``Fabrication of vapor and gas sensors using films of aligned
  {CN$_x$} nanotubes''}, \newblock{}{Chem. Phys. Lett.} {}{\textbf{386}}(1-3),
  {}{137} ({}{Mar. 2004}), \newblock
  doi:\href{http://dx.doi.org/10.1016/j.cplett.2004.01.052}{10.1016/j.cplett.2%
004.01.052}.
\bibitem{rocha}
{}{A.~R. Rocha}, {}{M.~Rossi}, {}{A.~Fazzio}, and {}{A.~J.~R. da~Silva},
  \newblock{}{``Designing real nanotube-based gas sensors''}, \newblock{}{Phys.
  Rev. Lett.} {}{\textbf{100}}(17), {}{176803} ({}{May 2008}), \newblock
  doi:\href{http://dx.doi.org/10.1103/PhysRevLett.100.176803}{10.1103/PhysRevL%
ett.100.176803}.
\bibitem{Brahim}
{}{S.~Brahim}, {}{S.~Colbern}, {}{R.~Gump}, and {}{L.~Grigorian},
  \newblock{}{``Tailoring gas sensing properties of carbon nanotubes''},
  \newblock{}{J. Appl. Phys.} {}{\textbf{104}}(2), {}{024502} ({}{Jul. 2008}),
  \newblock doi:\href{http://dx.doi.org/10.1063/1.2956395}{10.1063/1.2956395}.
\bibitem{morgan}
{}{C.~Morgan}, {}{Z.~Alemipour}, and {}{M.~Baxendale}, \newblock{}{``Variable
  range hopping in oxygen-exposed single-wall carbon nanotube networks''},
  \newblock{}{Phys. Stat. Solidi A} {}{\textbf{205}}(6), {}{1394} ({}{May
  2008}), \newblock
  doi:\href{http://dx.doi.org/10.1002/pssa.200778113}{10.1002/pssa.200778113}.
\bibitem{cnt_networks}
{}{D.~J. Mowbray}, {}{C.~Morgan}, and {}{K.~S. Thygesen},
  \newblock{}{``Influence of {O}$_{\text{2}}$ and {N}$_{\text{2}}$ on the
  conductivity of carbon nanotube networks''}, \newblock{}{Phys. Rev. B}
  {}{\textbf{79}}(19), {}{195431} ({}{May 2009}), \newblock
  doi:\href{http://dx.doi.org/10.1103/PhysRevB.79.195431}{10.1103/PhysRevB.79.%
195431}.
\bibitem{Fagan}
{}{S.~B. Fagan}, {}{R.~Mota}, {}{A.~J.~R. da~Silva}, and {}{A.~Fazzio},
  \newblock{}{``\emph{Ab initio} study of an iron atom interacting with
  single-wall carbon nanotubes''}, \newblock{}{Phys. Rev. B}
  {}{\textbf{67}}(20), {}{205414} ({}{May 2003}), \newblock
  doi:\href{http://dx.doi.org/10.1103/PhysRevB.67.205414}{10.1103/PhysRevB.67.%
205414}.
\bibitem{Yagi}
{}{Y.~Yagi}, {}{T.~M. Briere}, {}{M.~H.~F. Sluiter}, {}{V.~Kumar}, {}{A.~A.
  Farajian}, and {}{Y.~Kawazoe}, \newblock{}{``Stable geometries and magnetic
  properties of single-walled carbon nanotubes doped with $3d$ transition
  metals: A first-principles study''}, \newblock{}{Phys. Rev. B}
  {}{\textbf{69}}(7), {}{075414} ({}{Feb. 2004}), \newblock
  doi:\href{http://dx.doi.org/10.1103/PhysRevB.69.075414}{10.1103/PhysRevB.69.%
075414}.
\bibitem{Yang}
{}{S.~H. Yang}, {}{W.~H. Shin}, {}{J.~W. Lee}, {}{S.~Y. Kim}, {}{S.~I. Woo},
  and {}{J.~K. Kang}, \newblock{}{``Interaction of a transition metal atom with
  intrinsic defects in single-walled carbon nanotubes''}, \newblock{}{J. Phys.
  Chem. B} {}{\textbf{110}}(28), {}{13941} ({}{Jun. 2006}), \newblock
  doi:\href{http://dx.doi.org/10.1021/jp061895q}{10.1021/jp061895q}.
\bibitem{Chan}
{}{K.~T. Chan}, {}{J.~B. Neaton}, and {}{M.~L. Cohen},
  \newblock{}{``First-principles study of metal adatom adsorption on
  graphene''}, \newblock{}{Phys. Rev. B} {}{\textbf{77}}(23), {}{235430}
  ({}{Jun. 2008}), \newblock
  doi:\href{http://dx.doi.org/10.1103/PhysRevB.77.235430}{10.1103/PhysRevB.77.%
235430}.
\bibitem{Yeung}
{}{C.~S. Yeung}, {}{L.~V. Liu}, and {}{Y.~A. Wang}, \newblock{}{``Adsorption of
  small gas molecules onto {Pt}-doped single-walled carbon nanotubes''},
  \newblock{}{J. Phys. Chem. C} {}{\textbf{112}}(19), {}{7401} ({}{Apr. 2008}),
  \newblock doi:\href{http://dx.doi.org/10.1021/jp0753981}{10.1021/jp0753981}.
\bibitem{Vo}
{}{T.~Vo}, {}{Y.-D. Wu}, {}{R.~Car}, and {}{M.~Robert},
  \newblock{}{``Structures, interactions, and ferromagnetism of {Fe}-carbon
  nanotube systems''}, \newblock{}{J. Phys. Chem. C} {}{\textbf{112}}(22),
  {}{400} ({}{May 2008}), \newblock
  doi:\href{http://dx.doi.org/10.1021/jp0761968}{10.1021/jp0761968}.
\bibitem{Furst}
{}{J.~A. F\"{u}rst}, {}{M.~Brandbyge}, {}{A.-P. Jauho}, and {}{K.~Stokbro},
  \newblock{}{``\emph{Ab initio} study of spin-dependent transport in carbon
  nanotubes with iron and vanadium adatoms''}, \newblock{}{Phys. Rev. B}
  {}{\textbf{78}}(19), {}{195405} ({}{Nov. 2008}), \newblock
  doi:\href{http://dx.doi.org/10.1103/PhysRevB.78.195405}{10.1103/PhysRevB.78.%
195405}.
\bibitem{Juanma}
{}{J.~M. Garc\'{\i}a-Lastra}, {}{K.~S. Thygesen}, {}{M.~Strange}, and
  {}{\'{A}ngel Rubio}, \newblock{}{``Conductance of sidewall-functionalized
  carbon nanotubes: Universal dependence on adsorption sites''},
  \newblock{}{Phys. Rev. Lett.} {}{\textbf{101}}(23), {}{236806} ({}{Dec.
  2008}), \newblock
  doi:\href{http://dx.doi.org/10.1103/PhysRevLett.101.236806}{10.1103/PhysRevL%
ett.101.236806}.
\bibitem{Krasheninnikov}
{}{A.~V. Krasheninnikov}, {}{P.~O. Lehtinen}, {}{A.~S. Foster},
  {}{P.~Pyykk\"{o}}, and {}{R.~M. Nieminen}, \newblock{}{``Embedding
  transition-metal atoms in graphene: Structure, bonding, and magnetism''},
  \newblock{}{Phys. Rev. Lett.} {}{\textbf{102}}(12), {}{126807} ({}{Mar.
  2009}), \newblock
  doi:\href{http://dx.doi.org/10.1103/PhysRevLett.102.126807}{10.1103/PhysRevL%
ett.102.126807}.
\bibitem{Kramberger07PRB}
{}{C.~Kramberger}, {}{H.~Rauf}, {}{H.~Shiozawa}, {}{M.~Knupfer},
  {}{B.~Buchner}, {}{T.~Pichler}, {}{D.~Batchelor}, and {}{H.~Kataura},
  \newblock{}{``Unraveling van hove singularities in x-ray absorption response
  of single-wall carbon nanotubes''}, \newblock{}{Phys. Rev. B}
  {}{\textbf{75}}(23), {}{235437} ({}{Jun. 2007}), \newblock
  doi:\href{http://dx.doi.org/10.1103/PhysRevB.75.235437}{10.1103/PhysRevB.75.%
235437}.
\bibitem{Ayala09PRB}
{}{P.~Ayala}, {}{Y.~Miyata}, {}{K.~De~Blauwe}, {}{H.~Shiozawa}, {}{Y.~Feng},
  {}{K.~Yanagi}, {}{C.~Kramberger}, {}{S.~R.~P. Silva}, {}{R.~Follath},
  {}{H.~Kataura}, and {}{T.~Pichler}, \newblock{}{``Disentanglement of the
  electronic properties of metallicity-selected single-walled carbon
  nanotubes''}, \newblock{}{Phys. Rev. B} {}{\textbf{80}}(20), {}{205427}
  ({}{Nov. 2009}), \newblock
  doi:\href{http://dx.doi.org/10.1103/PhysRevB.80.205427}{10.1103/PhysRevB.80.%
205427}.
\bibitem{DeBlauwe2010}
{}{K.~De~Blauwe}, {}{D.~J. Mowbray}, {}{Y.~Miyata}, {}{P.~Ayala}, {}{A.~Rubio},
  {}{P.~Hoffman}, {}{H.~Kataura}, and {}{T.~Pichler} ({}{2010}), (submitted).
\bibitem{Zheng03NM}
{}{M.~Zheng}, {}{A.~Jagota}, {}{E.~D. Semke}, {}{B.~A. Diner}, {}{R.~S.
  Mclean}, {}{S.~R. Lustig}, {}{R.~E. Richardson}, and {}{N.~G. Tassi},
  \newblock{}{``Dna-assisted dispersion and separation of carbon nanotubes''},
  \newblock{}{Nature Materials} {}{\textbf{2}}(5), {}{338} ({}{May 2003}),
  \newblock doi:\href{http://dx.doi.org/10.1038/nmat877}{10.1038/nmat877}.
\bibitem{Zheng03S}
{}{M.~Zheng}, {}{A.~Jagota}, {}{M.~S. Strano}, {}{A.~P. Santos}, {}{P.~Barone},
  {}{S.~G. Chou}, {}{B.~A. Diner}, {}{M.~S. Dresselhaus}, {}{R.~S. McLean},
  {}{G.~B. Onoa}, {}{G.~G. Samsonidze}, {}{E.~D. Semke}, {}{M.~Usrey}, and
  {}{D.~J. Walls}, \newblock{}{``Structure-based carbon nanotube sorting by
  sequence-dependent dna assembly''}, \newblock{}{Science}
  {}{\textbf{302}}(5650), {}{1545} ({}{Nov. 2003}), \newblock
  doi:\href{http://dx.doi.org/10.1126/science.1091911}{10.1126/science.1091911%
}.
\bibitem{TU09N}
{}{X.~M. Tu}, {}{S.~Manohar}, {}{A.~Jagota}, and {}{M.~Zheng},
  \newblock{}{``Dna sequence motifs for structure-specific recognition and
  separation of carbon nanotubes''}, \newblock{}{Nature}
  {}{\textbf{460}}(7252), {}{250} ({}{Jul. 2009}), \newblock
  doi:\href{http://dx.doi.org/10.1038/nature08116}{10.1038/nature08116}.
\bibitem{Li07JOTACS}
{}{X.~L. Li}, {}{X.~M. Tu}, {}{S.~Zaric}, {}{K.~Welsher}, {}{W.~S. Seo},
  {}{W.~Zhao}, and {}{H.~J. Dai}, \newblock{}{``Selective synthesis combined
  with chemical separation of single-walled carbon nanotubes for chirality
  selection''}, \newblock{}{J. Amer. Chem. Soc.} {}{\textbf{129}}(51),
  {}{15770} ({}{Dec. 2007}), \newblock
  doi:\href{http://dx.doi.org/10.1021/ja077886s}{10.1021/ja077886s}.
\bibitem{Fagan07JOTACS}
{}{J.~A. Fagan}, {}{J.~R. Simpson}, {}{B.~J. Bauer}, {}{S.~H.~D. Lacerda},
  {}{M.~L. Becker}, {}{J.~Chun}, {}{K.~B. Migler}, {}{A.~R.~H. Walker}, and
  {}{E.~K. Hobbie}, \newblock{}{``Length-dependent optical effects in
  single-wall carbon nanotubes''}, \newblock{}{J. Amer. Chem. Soc.}
  {}{\textbf{129}}(34), {}{10607} ({}{Aug. 2007}), \newblock
  doi:\href{http://dx.doi.org/10.1021/ja073115c}{10.1021/ja073115c}.
\bibitem{Arnold06NN}
{}{M.~S. Arnold}, {}{A.~A. Green}, {}{J.~F. Hulvat}, {}{S.~I. Stupp}, and
  {}{M.~C. Hersam}, \newblock{}{``Sorting carbon nanotubes by electronic
  structure using density differentiation''}, \newblock{}{Nature
  Nanotechnology} {}{\textbf{1}}(1), {}{60} ({}{Oct. 2006}), \newblock
  doi:\href{http://dx.doi.org/10.1038/nnano.2006.52}{10.1038/nnano.2006.52}.
\bibitem{Bachtold}
{}{B.~Lassagne}, {}{D.~Garcia-Sanchez}, {}{A.~Aguasca}, and {}{A.~Bachtold},
  \newblock{}{``Ultrasensitive mass sensing with a nanotube electromechanical
  resonator''}, \newblock{}{Nano Lett.} {}{\textbf{8}}(11), {}{3735} ({}{Oct.
  2008}), \newblock
  doi:\href{http://dx.doi.org/10.1021/nl801982v}{10.1021/nl801982v}.
\bibitem{ZettlMassSensor}
{}{K.~Jensen}, {}{K.~Kim}, and {}{A.~Zettl}, \newblock{}{``An atomic-resolution
  nanomechanical mass sensor''}, \newblock{}{Nature Nanotech.} {}{\textbf{3}},
  {}{533} ({}{Sep. 2008}), \newblock
  doi:\href{http://dx.doi.org/10.1038/nnano.2008.200}{10.1038/nnano.2008.200}.
\bibitem{Chuanhong}
{}{C.~Jin}, {}{H.~Lan}, {}{K.~Suenaga}, {}{L.~Peng}, and {}{S.~Iijima},
  \newblock{}{``Metal atom catalyzed enlargement of fullerenes''},
  \newblock{}{Phys. Rev. Lett.} {}{\textbf{101}}(17), {}{176102} ({}{Oct.
  2008}), \newblock
  doi:\href{http://dx.doi.org/10.1103/PhysRevLett.101.176102}{10.1103/PhysRevL%
ett.101.176102}.
\bibitem{Juanma2}
{}{J.~M. {Garc{\'{\i}}a-Lastra}}, {}{{D. J. Mowbray}}, {}{K.~S. Thygesen},
  {}{{\'{A}}.~Rubio}, and {}{K.~W. Jacobsen} ({}{2010}), \newblock
  arXiv:\href{http://arXiv.org/abs/1001.2538v1}{1001.2538v1},
  [cond-mat:mes-hall] (submitted).
\bibitem{PBE}
{}{J.~P. Perdew}, {}{K.~Burke}, and {}{M.~Ernzerhof}, \newblock{}{``Generalized
  gradient approximation made simple''}, \newblock{}{Phys. Rev. Lett.}
  {}{\textbf{77}}(18), {}{3865} ({}{Oct. 1996}), \newblock
  doi:\href{http://dx.doi.org/10.1103/PhysRevLett.77.3865}{10.1103/PhysRevLett%
.77.3865}.
\bibitem{GPAW}
{}{J.~J. Mortensen}, {}{L.~B. Hansen}, and {}{K.~W. Jacobsen},
  \newblock{}{``Real-space grid implementation of the projector augmented wave
  method''}, \newblock{}{Phys. Rev. B} {}{\textbf{71}}(3), {}{035109} ({}{Jan.
  2005}), \newblock
  doi:\href{http://dx.doi.org/10.1103/PhysRevB.71.035109}{10.1103/PhysRevB.71.%
035109}.
\bibitem{benchmark}
{}{M.~Strange}, {}{I.~S. Kristensen}, {}{K.~S. Thygesen}, and {}{K.~W.
  Jacobsen}, \newblock{}{``Benchmark density functional theory calculations for
  nanoscale conductance''}, \newblock{}{J. Chem. Phys.} {}{\textbf{128}}(11),
  {}{114714} ({}{Mar. 2008}), \newblock
  doi:\href{http://dx.doi.org/10.1063/1.2839275}{10.1063/1.2839275}.
\bibitem{SIESTA}
{}{J.~M. Soler}, {}{E.~Artacho}, {}{J.~D. Gale}, {}{A.~Garcia},
  {}{J.~Junquera}, {}{P.~Ordej\'{o}n}, and {}{D.~S\'{a}nchez-Portal},
  \newblock{}{``The {SIESTA} method for \emph{ab initio} order-$n$ materials
  simulation''}, \newblock{}{J. Phys.: Condens. Matter} {}{\textbf{14}}(11),
  {}{2745} ({}{Mar. 2002}), \newblock
  doi:\href{http://dx.doi.org/10.1088/0953-8984/14/11/302}{10.1088/0953-8984/1%
4/11/302}.
\bibitem{CRCHandbook}
{}{D.~Lide}, \newblock{}{\emph{Handbook of Chemistry and Physics}}, 87th ed.
  ({}{CRC-Press}, {}{2006--2007}).
\end{thebibliography}

\end{document}